\documentclass[sigconf,natbib=true,anonymous=false]{acmart}
\AtBeginDocument{%
  \providecommand\BibTeX{{%
    \normalfont B\kern-0.5em{\scshape i\kern-0.25em b}\kern-0.8em\TeX}}}

\setcopyright{acmlicensed}
\copyrightyear{2018}
\acmYear{2018}
\acmDOI{XXXXXXX.XXXXXXX}

\acmConference{Under Review}{2024}{}
\acmBooktitle{Woodstock '18: ACM Symposium on Neural Gaze Detection,
 June 03--05, 2018, Woodstock, NY} 
\acmISBN{978-1-4503-XXXX-X/18/06}

\usepackage{hyperref}       
\usepackage{url}            
\usepackage{booktabs}       
\usepackage{amsfonts}       
\usepackage{nicefrac}       
\usepackage{microtype}      

\usepackage{microtype}
\usepackage{amsmath}
\usepackage{algpseudocode}
\usepackage{algorithm}
\usepackage{multirow}
\usepackage{subcaption}
\usepackage{bm}
\usepackage{stfloats}
\usepackage{arydshln}
\usepackage{graphicx}  
\usepackage{comment}
\usepackage{enumitem}

\newcommand{\eg}{\textit{e.g.}}
\renewcommand\footnotetextcopyrightpermission[1]{}
\begin{document}

\title{\textsc{ReFIT}: Relevance Feedback from a Reranker during Inference}
\pagestyle{plain}
\author{\textbf{Revanth Gangi Reddy}$^1$ \hspace{0.2em} \textbf{Pradeep Dasigi}$^2$ \hspace{0.2em} \textbf{Md Arafat Sultan}$^3$ \hspace{0.2em} \textbf{Arman Cohan}$^{2,4}$ \\ \textbf{Avirup Sil}$^3$ \hspace{0.2em} \textbf{Heng Ji}$^1$ \hspace{0.2em} \textbf{Hannaneh Hajishirzi}$^{2,5}$ \\
$^1$University of Illinois at Urbana-Champaign \hspace{0.5em}  $^2$Allen Institute for AI \\ $^3$IBM Research AI \hspace{0.5em} $^4$Yale University \hspace{0.2em} $^5$University of Washington\\
  \texttt{\{revanth3,hengji\}@illinois.edu} \hspace{0.4em} \texttt{\{pradeepd,armanc,hannah\}@allenai.org} \\
  \texttt{arafat.sultan@ibm.com}\hspace{0.4em} \texttt{avi@us.ibm.com}
  }

\renewcommand{\shortauthors}{}

\begin{abstract}

Retrieve-and-rerank is a prevalent framework in neural information retrieval, wherein a bi-encoder network initially retrieves a pre-defined number of candidates (\eg{}, $K$=100), which are then reranked by a more powerful cross-encoder model. While the reranker often yields improved candidate scores compared to the retriever, its scope is confined to only the top $K$ retrieved candidates. As a result, the reranker cannot improve retrieval performance in terms of Recall@K. In this work, we propose to leverage the reranker to improve recall by making it provide relevance feedback to the retriever at \textit{inference-time}. Specifically, given a test instance during inference, we distill the reranker's predictions for that instance into the retriever's query representation using a lightweight update mechanism. The aim of the distillation loss is to align the retriever's candidate scores more closely with those produced by the reranker. The algorithm then proceeds by executing a second retrieval step using the updated query vector. We empirically demonstrate that this method, applicable to various retrieve-and-rerank frameworks, substantially enhances the retrieval recall across multiple domains, languages, and modalities.

\end{abstract}

\maketitle

\section{Introduction}


Information Retrieval (IR) involves retrieving a set of candidates from a large document collection
given a user query. The retrieved candidates may be further reranked to bring the most relevant ones to the top, constituting a typical retrieve-and-rerank (R\&R) framework \cite{wang2018evidence, hu2019retrieve}.
Reranking generally improves the ranks of relevant candidates among those retrieved, thus improving on metrics such as Mean Reciprocal Rank (MRR) \cite{Craswell2009} and Normalized Discounted Cumulative Gain (nDCG) \cite{jarvelin2002cumulated}, which assign better scores when relevant results are ranked higher. 
However, retrieval metrics like Recall@K, which mainly evaluate the presence of relevant candidates in the top $K$ retrieved results, remain unaffected.
Increasing Recall@K
can be 
key, especially when the retrieved results are used in downstream knowledge-intensive tasks \cite{petroni2021kilt} such as open-domain question answering \cite{chen2017reading, chen2020open, gangi2021synthetic}, fact-checking \cite{thorne2018fever}, entity linking \cite{hoffart2011robust,sil2013re,sil2018neural} and dialog generation \cite{dinan2018wizard, komeili2022internet}.
\begin{figure}[t]
    \centering
    \includegraphics[width=1.0\linewidth]{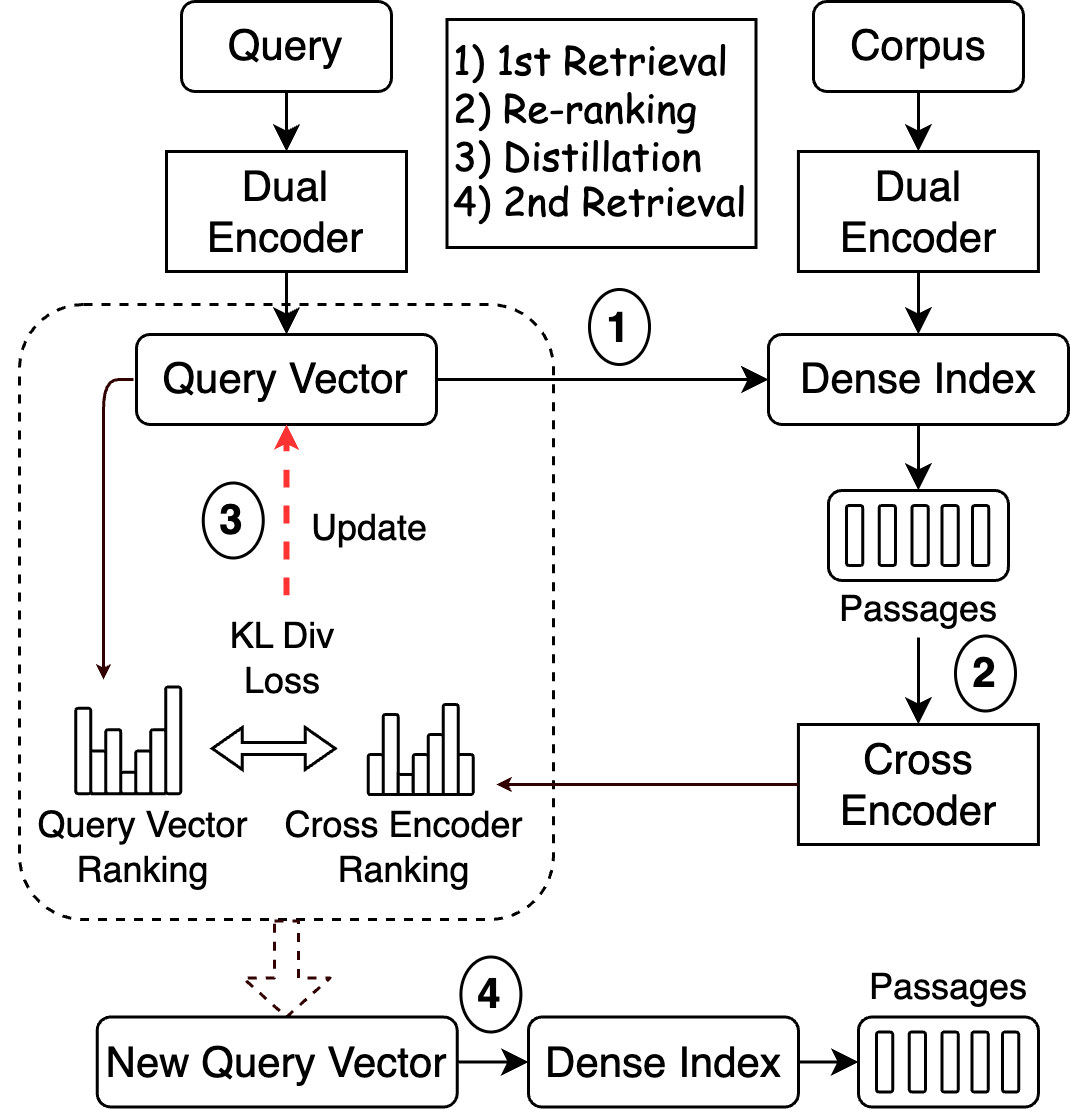}
    \caption{\textsc{ReFIT}: The proposed method for reranker relevance feedback. We introduce an inference-time distillation process (step 3) into the traditional retrieve-and-rerank framework (steps 1 and 2) to compute a new query vector, which improves recall when used for a second retrieval step (step 4).}
    \label{fig:overall_framework}
\end{figure}

Most existing neural IR methods use a dual-encoder retriever \cite{karpukhin2020dense, khattab2020colbert} and a subsequent cross-encoder reranker \cite{nogueira2019passage}. 
Dual-encoder\footnote{We use the terms bi-encoder and dual-encoder interchangeably in this paper.} models leverage separate query and passage encoders and perform a late interaction between the query and passage output representations. This enables them to perform inference at scale as passage representations can be pre-computed. Cross-encoder models, on the other hand, accept the query and the passage together as input, leaving out scope for pre-computation. The cross-encoder typically provides better ranking than the dual-encoder---thanks to its more elaborate computation of query-passage similarity informed by cross-attention---but is limited to seeing only the retrieved candidates in an 
R\&R
framework.




Since the more sophisticated reranker often generalizes better at passage scoring than the simpler, but more efficient retriever, here we propose to use relevance feedback from the former to improve the quality of query representations for the latter directly \textit{at inference}.
Concretely, after the R\&R pipeline is invoked for a test instance, we update the retriever's corresponding query vector by minimizing a distillation loss that brings its score distribution over the retrieved passages closer to that of the reranker.
The new query vector is then used to retrieve documents for the second time.
This process effectively teaches the retriever how to rank passages like the reranker---a stronger model---for the given test instance.
Our approach, \textsc{ReFIT}\footnote{\textsc{ReFIT} stands for \textbf{Re}ranker \textbf{F}eedback at \textbf{I}nference \textbf{T}ime}, is lightweight as only the output query vectors (and no model parameters) are updated, ensuring comparable inference-time latency when incorporated into the 
R\&R
framework. 
Figure \ref{fig:overall_framework} shows a schematic diagram of our approach, which introduces a distillation and a second retrieval step into the R\&R framework.
By operating exclusively in the representation space---as we only update the query vectors---our framework yields a parameter-free and architecture-agnostic solution, thereby providing flexibility along important application dimensions, e.g., the language, domain, and modality of retrieval. 
We empirically demonstrate this effect by showing improvements in retrieval on multiple English domains, across 26 languages in multilingual and cross-lingual settings, and in different modalities such as text and video retrieval.


Our main contributions are as follows:
\begin{itemize}
    \item We propose \textsc{ReFIT}, an inference-time mechanism to improve the recall of retrieval in IR using relevance feedback from a reranker.
    \item Empirically, \textsc{ReFIT} improves retrieval performance in multi-domain, multilingual, cross-lingual and multi-modal evaluation.
    \item The proposed distillation step
    is fast, considerably increasing recall without any loss in ranking performance over a standard R\&R pipeline with comparable latency.
\end{itemize}

\section{Related Work}
\label{sec:related-work}
\noindent \textbf{Pseudo-relevance feedback:} Our method has similarities with 
Pseudo-Relevance Feedback (PRF) \cite{rocchio1971relevance, lv2009adaptive, li2022does} in IR: \citet{bendersky2011parameterized, xu2017quary} use the retrieved documents to improve sparse approaches via query expansion or query term reweighting, \citet{li2018nprf, zheng2020bert} score similarity between a target document and a top-ranked feedback document, while \citet{yu2021improving} train a separate query encoder that computes a new query embedding using the retrieved documents as additional input. In contrast, our approach does not require customized training feedback models or availability of explicit feedback data, as we improve the query vector by directly distilling from the reranker's output within an R\&R framework. 

Further, previous approaches to PRF have been dependent on the choice of retriever architecture and language; \citet{yu2021improving}'s PRF model is tied to the retriever used, \citet{chandradevan2022learning} explore cross-lingual relevance feedback, but require feedback documents in target language and thereby could only apply to three languages, while \citet{li2022interpolate} explore interpolating relevance feedback between dense and sparse approaches.
On the other hand, our approach is independent of the choice of the retriever and reranker architecture, and can be used for neural retrieval in any domain, language or modality. \\

\noindent \textbf{Distillation in Neural IR:} Existing approaches primarily leverage reranker feedback \textit{during training} of the dual-encoder retriever, to sample better negatives \cite{qu2021rocketqa}, for standard knowledge distillation of the cross-attention scores \cite{izacard2020distilling}, to train smaller and more efficient rankers by distilling larger models \cite{hofstatter2020improving}, or to align the geometry of dual-encoder embeddings with that from cross-encoders \cite{wang2021enhancing}. Instead, we leverage distillation at inference time, updating only the query representation to replicate the cross-encoder’s scores for the corresponding test instance.
A key implication of this design choice is that unlike existing methods, we keep the retriever parameters unchanged, meaning \textsc{ReFIT} can be incorporated out-of-the-box into any neural R\&R framework. In contrast, extending training-time distillation to new languages or modalities would require re-training the bi-encoder.

More recently, \textsc{TouR}~\cite{sung2023optimizing} has proposed test-time optimization of query representations with two variants: \textsc{TouR}$_{\text{hard}}$ and  \textsc{TouR}$_{\text{soft}}$. 
\textsc{TouR}$_{\text{hard}}$ optimizes the marginal likelihood of a small set of (pseudo) positive contexts.
\textsc{ReFIT} shares similarities with \textsc{TouR}$_{\text{soft}}$, which uses the normalized scores of a cross-encoder over the retrieved results as soft labels.
Crucially, \textsc{TouR} relies on multiple iterations of relevance feedback via distillation, where each iteration runs until the top-1 retrieval result has the highest reranker score (in \textsc{TouR}$_{\text{soft}}$) or is a pseudo-positive (in \textsc{TouR}$_{\text{hard}}$).
This makes inference highly computationally expensive, as each additional iteration involves labeling top-$K$ retrieval results with a reranker and then retrieving again.
\textsc{ReFIT} improves efficiency over \textsc{TouR} by requiring only a single iteration of feedback that simply updates the query vector for longer, foregoing additional retrieval and reranking steps. More specifics on the inference process of the two methods can be found in \S{\ref{sec:tour_comparison}}.
\textsc{TouR} was evaluated only on English phrase and passage retrieval tasks, while we demonstrate \textsc{ReFIT}'s effectiveness in multidomain, multilingual and multimodal settings, with an empirical comparison with \textsc{TouR} in \S{\ref{sec:tour_comparison}}.

\section{Method}

Here we discuss the standard retrieve-and-rerank (R\&R) framework for IR (\S{\ref{sec:retrieve_and_rerank}}) and  
how our proposal fits into it (\S{\ref{sec:cross_encoder_feedback}}). While our approach can be applied to any R\&R framework, we consider a text-based retriever and reranker for simplicity while elaborating our method. A multi-modal R\&R framework is described in \S\ref{sec:multimodal_results}.

\subsection{Retrieve-and-Rerank}
\label{sec:retrieve_and_rerank}
R\&R for IR consists of a first-stage retriever and a second-stage reranker. Modern neural approaches typically use a dual-encoder model as the retriever and a cross-encoder for reranking.  

\paragraph{\textbf{The Retriever}:} The dual-encoder retriever model is based on a Siamese neural network \cite{chicco2021siamese}, containing separate Bert-based \cite{devlin2019bert} encoders $E_Q(.)$ and $E_P(.)$ for the query and the passage, respectively.
Given a query $q$ and a passage $p$, a separate representation is obtained for each, such as the \textsc{cls} output or a pooled representation of the individual token outputs from $E_Q(q)$ and $E_P(p)$. 
The question-passage similarity $sim(q,p)$ is computed as the dot product of their corresponding representations:
\begin{equation}
    Q_q = Pool(E_Q(q))
\end{equation}
\begin{equation}
    P_p = Pool(E_P(p))
\end{equation}
\begin{equation}\label{eq:sim}
   sim(q,p) = S(Q_q,P_p) = Q_q^TP_p
\end{equation}

Since Eq.~\ref{eq:sim} is decomposable, the representations of all passages in the retrieval corpus can be pre-computed and stored in a dense index \cite{johnson2019billion}. During inference, given a new query, the top $K$ most relevant passages are retrieved from the index via approximate nearest-neighbor search.

\paragraph{\textbf{The Reranker}:} The cross-encoder reranker model uses a Bert-based encoder $E_R(.)$, which takes the query $q$ and a corresponding retrieved passage $p$ together as input and outputs a similarity score. 
A feed-forward layer $F$ is used on top of the \textsc{cls} output from $E_R(.)$ to compute a single logit, which is used as the final reranker score $R(q,p)$. The top $K$ retrieved passages are then ranked based on their corresponding reranker scores.

\begin{equation}
   R(q,p) = F(CLS(E_R(q,p))
\end{equation}


\begin{algorithm}[t]
\caption{\textsc{\textbf{ReFIT}}}
\label{alg4}
\begin{flushleft}
\textbf{Input}: Query $q$ and its representation $Q_q$, retrieved passages $P$ and their representations $\hat{P}$.\newline
\textbf{Output}: Updated query representation $Q_{q,n}$
\end{flushleft}
\begin{algorithmic}[1]
    \State Initialize query vector $Q_{q,0}$ = $Q_q$
    \State Compute reranker distribution $D_{CE}(q,P)$ (Eq.~\ref{eq:d-ce})
    \For{\textit{i in 0 to n}}
        \State Compute retriever distribution $D_{Q_{q,i}}(\hat{P})$ (Eq.~\ref{eq:d-q})
        \State Compute loss $\mathcal{L}$ (Eq.~\ref{eq:loss})
        \State Update $Q_{q,i+1} = Q_{q,i} - \alpha \frac{\partial}{\partial Q_{q,i}}\mathcal{L}$
    \EndFor
    \State return $Q_{q,n}$
\end{algorithmic}
\end{algorithm}

\subsection{Reranker Relevance Feedback}
\label{sec:cross_encoder_feedback}
The main idea underlying our proposal is to compute an improved query representation for the retriever using feedback from the more powerful reranker.
More specifically, we perform a lightweight inference-time distillation of the reranker's knowledge into a new query vector.

Given an input query $q$ during inference, we use the following output provided by the R\&R pipeline:
\begin{itemize}
   \item Query representation $Q_q$ from the retriever.
    \item Retrieved passages $P = \{p_1, p_2,  ..., p_K\}$ and their representations $\hat{P} = [P_{p_1}, P_{p_1},  ..., P_{p_K}]$ from the retriever. 
    \item The reranking scores $R(q,P) = [R(q,p_1),..., R(q,p_K)]$.
\end{itemize}
Note that $\hat{P}$ above is directly obtained from the passage index and is not computed during inference.

The proposed reranker feedback mechanism begins with using the reranking scores $R(q,P)$ to compute a cross-encoder ranking distribution $D_{CE}(q,P)$ over passages $P$ as follows:

\begin{equation}
D_{CE}(q,P)=\mathrm{softmax}([R(q,p_1), ..., R(q,p_K)])
     \label{eq:d-ce}
\end{equation} 

The query and passage representations from the retriever are used to compute a similar distribution $D_{Q_q}(\hat{P})$ over $P$:

\begin{equation}
    D_{Q_q}(\hat{P}) = \mathrm{softmax}([Q_q^TP_{p_1}, ..., Q_q^TP_{p_K}])
    \label{eq:d-q}
\end{equation}


Next, we compute the loss as the KL-divergence between the retriever and reranker distributions:

\begin{equation}
    \mathcal{L} = D_{KL}(D_{CE}(q,P) || D_{Q_q}(\hat{P}))
    \label{eq:loss}
\end{equation}

which is then used to update the query vector via gradient descent. The query vector update process is repeated for $n$ times, where $n$ is a hyper-parameter. 
A schematic description of the process can be found in Algorithm \ref{alg4}. 

Finally, the updated query vector $Q_{q,n}$ is used for a second-stage retrieval from the passage index.  
From dual-encoder retrieval with the updated $Q_{q,n}$, we aim to achieve better recall than with the initial $Q_q$, while obtaining a ranking performance that is comparable with that of the reranker.

\section{Experimental Setup}
\label{sec:exp_setup}
\subsection{Distillation Process} We observe that the output scores from the dual-encoder and the cross-encoder models are not bounded to specific intervals. Hence, we do min-max normalization separately on the query vector's scores $Q^T_q\hat{P}$ (from the dual-encoder) and the cross encoder's scores $R(q,P)$ to bring the two scoring distributions closer. Further, the cross-encoder tends to have peaky scoring distributions, hence we use a temperature $T$ (= 2 after hyperparameter-tuning) while computing the softmax $D_{CE}(q,P)$ over the cross-encoder scores. After tuning on the MS Marco dev set, we set the number of updates $n$=100 with learning rate $\alpha$=0.005.

\subsection{Rerank Baseline}
\begin{table}
\setlength{\tabcolsep}{1.0em}
        \centering
        \def\arraystretch{1.2}
        \begin{tabular}{c|cc|c}
        \hline
        \multirow{2}{*}{\textbf{Step (Device)}} & \multicolumn{2}{c}{\textbf{Retrieve \& Rerank}} & \multirow{2}{*}{\parbox{3em}{\centering\textbf{\textsc{ReFIT} (K=100)}}}\\
        \cline{2-3}
        & \multicolumn{1}{|c}{\textbf{K=100}} & \multicolumn{1}{c|}{\textbf{K=125}} & \\
        \hline
        1st Retrieval (CPU) & \multicolumn{1}{c}{40ms} & \multicolumn{1}{c|}{40ms} & \multicolumn{1}{c}{40ms} \\
        \hdashline
        Rerank (CPU) & \multicolumn{1}{c}{1540ms} & \multicolumn{1}{c|}{1925ms} & \multicolumn{1}{c}{1540ms} \\
        Rerank (GPU) & \multicolumn{1}{c}{360ms} & \multicolumn{1}{c|}{450ms} & \multicolumn{1}{c}{360ms} \\
        \hdashline
        Distillation (CPU) & \multicolumn{1}{c}{-} & \multicolumn{1}{c|}{-}&  \multicolumn{1}{c}{30ms}\\
        2nd Retrieval (CPU) & \multicolumn{1}{c}{-} & \multicolumn{1}{c|}{-} & \multicolumn{1}{c}{40ms}\\
        \hline
        Total (CPU) & \multicolumn{1}{c}{1580ms}&  \multicolumn{1}{c|}{1965ms} & \multicolumn{1}{c}{1650ms}\\
        Total (GPU) & \multicolumn{1}{c}{400ms}&  \multicolumn{1}{c|}{490ms} & \multicolumn{1}{c}{470ms}\\
        \hline
        \end{tabular}
        \caption{Comparison of inference times (in milliseconds) for different approaches, utilizing both CPU-only and GPU configurations (when reranking $K$ passages).}
        \label{tab:inf_times}
        \vspace{-1em}    
\end{table}

\textsc{ReFIT} introduces the additional overhead of distillation and a second retrieval step into the R\&R framework. We note that distillation latency (in Algorithm \ref{alg4}) in linear in the number of updates $n$.
Table \ref{tab:inf_times} 
compares the inference latency of our method with that of standard
R\&R, assuming $K$=100 passages are to be reranked and $n$=100 updates are used during distillation. We highlight that our distillation process is lightweight and takes just 30ms on a CPU. We see that the additional distillation and retrieval steps increase the latency of inference by roughly 17.5\% when using a GPU (or 4.4\% for CPU);\footnote{24-core AMD EPYC 7352 CPU and 80GB A100 GPU.}
in that same amount of time, vanilla R\&R can process a total of 125 passages on the GPU (see Table~\ref{tab:inf_times}), to potentially increase Recall@100.
In view of this observation, and for fair comparison, we evaluate against a \mbox{\textit{Rerank}} baseline that is allowed to retrieve and rerank 125 passages. 
We note that both \textsc{ReFIT} and the \textit{Rerank} baseline use the same retriever and reranker, and are evaluated on Recall@100. 

\subsection{Retriever and Reranker} 
\label{sec:retriever_and_reranker}
We use Contriever \cite{izacard2021unsupervised} as the underlying retriever (unless otherwise mentioned), which has been pre-trained with an unsupervised contrastive learning objective on a large-scale collection of Wikipedia and CCNet 
documents. Contriever is a SOTA dual-encoder retriever that outperforms  traditional term-matching methods, BM25 
and recent dense approaches e.g. DPR \cite{karpukhin2020dense}, ANCE \cite{xiong2020approximate} and ColBERT \cite{khattab2020colbert}.
For retrieval in both English and other languages, we use the publicly available version of Contriever,
fine-tuned on MS MARCO \cite{nguyen2016ms}. Our English\footnote{\href{https://huggingface.co/cross-encoder/ms-marco-MiniLM-L-6-v2}{cross-encoder/ms-marco-MiniLM-L-6-v2}} and multilingual\footnote{\href{https://huggingface.co/cross-encoder/mmarco-mMiniLMv2-L12-H384-v1}{cross-encoder/mmarco-mMiniLMv2-L12-H384-v1}} rerankers are based on sentence transformers. 



\section{Results}
\label{sec:results}

\subsection{English Retrieval in Multiple Domains}
\label{sec:english_results}

\begin{table*}[t]
    \centering
    \small
    \setlength{\tabcolsep}{0.71em}
    \def\arraystretch{1.2}
    \begin{tabular}{c|cc:cccc|ccc}
    \hline
    \multicolumn{1}{c|}{\multirow{2}{*}{}} & \multicolumn{1}{c}{\multirow{2}{*}{\textbf{BM25}}} & \multicolumn{1}{c:}{\multirow{2}{*}{\textbf{ANCE}}} & \multicolumn{1}{c}{\multirow{2}{*}{\textbf{RocketQAv1}}} & \multicolumn{1}{c}{\multirow{2}{*}{\textbf{RocketQAv2}}} & \multicolumn{1}{c}{\textbf{RocketQAv1}}& \multicolumn{1}{c|}{\textbf{RocketQAv1}} & \multicolumn{1}{c}{\multirow{2}{*}{\textbf{Contriever}}} & \multicolumn{1}{c}{\textbf{Contriever}}& \multicolumn{1}{c}{\textbf{Contriever}}  \\
  & & & & & \textbf{+ Rerank} &  \textbf{+ \textsc{ReFIT}} & &  \textbf{+ Rerank} &  \textbf{+ \textsc{ReFIT}}  \\
    \hline
     MS MARCO & 65.8 &  85.2 & 88.4 & 88.7&	89.4& \textbf{90.0*}   & 89.1 & 89.9  & \textbf{90.5}* \\
       Trec-COVID  & 49.8  & 45.7  &48.5&	46.4&	52.0&	\textbf{52.9}   &  40.7 & 43.8  & \textbf{51.5}*\\
       NFCorpus   & 25.0   & 23.2 & 26.9	&25.9&	27.4	&\textbf{29.2*} & 30.0  & 29.5 & \textbf{31.9}*\\
       NQ  & 76.0 & 83.6 & 91.1	&89.8	&91.8	&\textbf{92.7*}  &  92.5 &  93.3  & \textbf{94.2}*\\
       HotpotQA  & 74.0 &  57.8   & 69.8	&67.7	&71.4	&\textbf{73.3*}   & 77.7 & 78.6  & \textbf{80.4}*\\
       FiQA  & 53.9   & 58.1 & 63.6	&61.2	&\textbf{64.3}	&63.8   & 65.6 &  \textbf{65.9}  & 65.6 \\
       DBPedia  & 39.8  &  31.9 & 45.7	&43.4	&47.6	&\textbf{50.2*}  &  54.1 & 56.0  & \textbf{57.3}*\\
       Scidocs & 35.6  &  26.9 & 31.8	&29.3	&33.1	&\textbf{35.5*}  &  37.8  & 38.3 & \textbf{40.1}*\\
       FEVER &  93.1   & 90.0 & 92.6	&92.5	&92.8	&\textbf{93.7*}   & 94.9 &  95.3  & \textbf{95.5}*\\
       Climate-FEVER  & 43.6   & 44.5  & 47.4	&48.7	&49.3&	\textbf{53.6*}    & 57.4  & 59.0  & \textbf{59.5}\\
       Scifact   & 90.8  & 81.6  & 88.1	&85.4	&89.0	&\textbf{89.9*}  & 94.7&  94.4 & \textbf{95.2*}\\
       \hline
       Average& 58.9  & 57.1 & 63.1	&61.7	&64.4	&\textbf{65.9*}  & 66.8 & 67.6  & \textbf{69.0}* \\
       \hline
    \end{tabular}
    \caption{Recall@100 (in \%) on the English BEIR benchmark. Performance of \textsc{ReFIT} is shown for different choices of underlying retrievers. RocketQAv2~\cite{ren2021rocketqav2} corresponds to a training-time distillation baseline. Improvements marked with * are statistically significant at $p<0.05$ as per paired t-test.}
    \label{tab:beir_nums}
\end{table*}

We evaluate English retrieval performance on the BEIR benchmark \cite{thakur2021beir}, comprising training and in-domain test instances from MS MARCO and out-of-domain evaluation data from a number of scientific, biomedical, financial, and Wikipedia-based retrieval datasets\footnote{We omit some datasets due to license \& versioning issues.}. 

Firstly, we compare our inference-time distillation approach against a training-time distillation method. We use RocketQAv1~\cite{qu-etal-2021-rocketqa} as the underlying retrieval model and RocketQAv2~\cite{ren-etal-2021-rocketqav2} as the retriever distilled at training time from the cross-encoder. We also compare with a \textit{Rerank} (K=125) baseline, which improves Recall@100 by reranking the top 125 passages (retrieved by RocketQAv1). Moreover, we also demonstrate the effectiveness of \textsc{ReFIT} with a different underlying retrieval model, in this case, Contriever. We refer to Appendix \S{\ref{sec:other_prf}} for discussion on experiments with other PRF baselines.

\begin{table}[t]
\small
    \centering
    \setlength{\tabcolsep}{0.75em}
    \def\arraystretch{1.2}
    \begin{tabular}{c|cc:ccc}
    \hline
    & \multicolumn{1}{c}{\parbox{2.8em}{\textbf{mBERT}}}  & \multicolumn{1}{c:}{\parbox{3.5em}{\textbf{XLM-R}}} & \multicolumn{1}{c}{\parbox{4em}{\textbf{Contriever}}} & \multicolumn{1}{c}{\parbox{3em}{\textbf{Rerank}}} & \multicolumn{1}{c}{\parbox{3em}{\textbf{\textsc{ReFIT}}}} \\
    \hline
    Arabic  & 81.1  & 79.9   & 88.7 &   89.5  &\textbf{90.9}* \\
    Bengali  &88.7   & 84.2   & 91.4&    91.4   & \textbf{95.9}* \\
    English  & 77.8    &  73.1  & 77.2&   78.7  &  \textbf{81.8}* \\
    Finnish   & 74.2   &  81.6  & 88.1 &    88.9  &\textbf{91.0}*\\
    Indonesian  & 81.0 & 87.4  & 89.8 &  90.5 &\textbf{93.7}*\\
    Japanese  & 76.1  &  70.9  & 81.7 & 82.5  &\textbf{85.2}*\\
    Korean  & 66.7  & 71.1  & 78.2&    \textbf{81.0}  &80.2\\
    Russian& 77.6  & 74.1    & 83.8 &   85.7  & \textbf{87.3}\\
    Swahili & 74.1 & 73.9   & 91.4  &  \textbf{92.0} &90.5\\
    Telugu & 89.5 & 91.2  & 96.6&    97.0 & \textbf{97.5}\\
    Thai& 57.8  &  89.5   & 90.5&    91.6  &\textbf{93.3}*\\
    \hline
    Average  & 76.8  & 79.7 & 87.0&   88.1 & \textbf{89.7}*\\
    \hline
    \end{tabular}
    \caption{Recall@100 (in \%) on the multilingual Mr.TyDi benchmark. Rerank and \textsc{ReFIT} use Contriever as the underlying retriever. Improvements with * are significant at $p<0.05$ according to the paired t-test.}
    \label{tab:mrtydi_nums}
    \vspace{-1em}
\end{table}

Table \ref{tab:beir_nums} shows Recall@100 results on the BEIR benchmark. Firstly, we see that \textsc{ReFIT} consistently outperforms all baselines. Next, RocketQAv2 shows improvement over RocketQAv1 on MS MARCO, which is the dataset used for training-time distillation of RocketQAv2. However, RocketQAv2's performance degrades on out-of-domain datasets from the BEIR benchmark. This is unsurprising\footnote{As a sanity check, we verified our reproduction of RocketQAv2 by comparing with other works (Table \ref{tab:rocketqa_sanity} in appendix)}, since the training-time distillation approach is limited to the bi-encoder seeing the cross-encoder’s relevance labels only in the source domain, i.e. the domain used for training (MS MARCO in this case). As a result, the training-time distillation approach may not generalize well to unseen domains (BEIR in this case). In contrast, \textsc{ReFIT} offers the key advantage of learning from target-domain pseudo labels provided by the reranker \textit{at inference}, which yields improved out-of-domain generalization. 



\subsection{Retrieval in More Languages}
\label{sec:multilingual_results}

\subsubsection{Multilingual Retrieval}
 We also evaluate on Mr. TyDi \cite{zhang2021mr}, a multilingual IR benchmark derived from TyDi QA \cite{clark2020tydi}, where given a question in one of 11 languages, the goal is to retrieve candidates from a pool of Wikipedia documents in the same language. Our underlying retriever is the multilingual version of Contriever. Other baseline retrieval models are mBERT and XLM-R \cite{conneau2020unsupervised}, in addition to the \textit{Rerank} (K=125) baseline. 
Table \ref{tab:mrtydi_nums} shows Recall@100 for the different systems on Mr.TyDi. Here again, \textsc{ReFIT} yields significant improvement over all baselines on most languages. 

\subsubsection{Cross-lingual Retrieval}
For our cross-lingual experiments, we used the MKQA benchmark \cite{longpre2021mkqa}. MKQA involves retrieving passages from the English Wikipedia corpus for questions that are posed in 26 different languages.
Following \cite{izacard2021unsupervised}, we discard unanswerable questions and questions with a yes/no answer or a long answer, leaving 6,619 queries per language in the final test set. 
Table \ref{tab:crossling_nums} compares Recall@100 of different models on MKQA. \textsc{ReFIT} again outperforms, leading the nearest baseline (\textit{Rerank}) by about 2 points on average, and with improvements on all 26 MKQA languages.

\begin{table*}[t]
    \centering
    \setlength{\tabcolsep}{0.8em}
    \def\arraystretch{1.2}
    \begin{tabular}{lcccccccccccccc}
    \hline
    & \textbf{avg} & \textbf{en} & \textbf{ar} & \textbf{fi} & \textbf{ja} & \textbf{ko} & \textbf{ru} & \textbf{es} & \textbf{sv} & \textbf{he} & \textbf{th} & \textbf{da} & \textbf{de} & \textbf{fr}\\ 
    \hline
    mBERT & 57.9 &74.2 & 44.0 & 51.7 & 55.7 &48.2 &57.4 &63.9 & 62.7 & 46.8 & 51.7 & 63.7 & 59.6 & 65.2\\
    XLM-R & 59.2 & 73.4 & 42.4 & 57.7& 53.1 &48.6 &58.5 &62.9 &67.5 &46.9 &61.5 &66.9 &60.9 &62.4\\
    \hdashline
    Contriever& 65.6 & 75.6 & 53.3 & 66.6 &60.4 &55.4 &64.7 &70.0 &70.8 &59.6 &63.5 &72.0 &66.6 &70.1\\
    Rerank & 66.4 & 76.0 & 54.5 & 67.5 & 61.5 & 56.7 & 65.8 & 70.5 & 71.6 & 60.8 & 64.9 & 72.7 & 67.5 & 70.6\\
    \textsc{ReFIT} & \textbf{68.2} & \textbf{76.6} & \textbf{58.0} & \textbf{68.8} & \textbf{64.7} & \textbf{59.3} & \textbf{68.4} & \textbf{72.5} & \textbf{73.1} & \textbf{62.9} & \textbf{66.5} & \textbf{74.1} & \textbf{70.1} & \textbf{72.5}\\
    \hline
    \hline
    &  & \textbf{it} & \textbf{nl} & \textbf{pl} & \textbf{pt} & \textbf{hu} & \textbf{vi} & \textbf{ms} & \textbf{km} & \textbf{no} & \textbf{tr} & \textbf{zh-cn} & \textbf{zh-hk} & \textbf{zh-tw} \\ 
    \hline
   mBERT & & 64.1 & 66.7 & 59.0 & 61.9 & 57.5 & 58.6 & 62.8 & 32.9 & 63.2 & 56.0 & 58.4 & 59.3 & 59.3\\
    XLM-R & & 58.1 & 66.4 & 61.0 & 62.0 & 60.1 & 62.4 & 66.1 & \textbf{46.6} & 65.9 & 60.6 & 55.8 & 55.5 & 55.7\\
    \hdashline
    Contriever & & 70.3 &71.4 &68.8 &68.5 &66.7 &67.8 &71.6 &37.8 &71.5 &68.7 &64.1 &64.5 &64.3 \\
    Rerank & & 70.8 & 72.0 & 69.9 & 69.3 & 67.5 & 68.7 & 72.0 & 38.6 & 72.3 & 69.3 & 65.1 & 65.4 & 65.2 \\
    \textsc{ReFIT} & & \textbf{72.4} & \textbf{73.6} & \textbf{71.1} & \textbf{71.5} & \textbf{68.9} & \textbf{70.5} & \textbf{73.3} & 39.9 &\textbf{73.3} & \textbf{70.7} & \textbf{67.5} & \textbf{67.4} & \textbf{66.9} \\
    \hline
    \hline
    \end{tabular}
    \caption{Recall@100 (in \%) on the cross-lingual MKQA benchmark.  Rerank and \textsc{ReFIT} use Contriever as the underlying retriever. All improvements are statistically significant at $p < 0.05$ as per paired t-test.}
    \label{tab:crossling_nums}
    \vspace{-0.5em}
\end{table*}

\subsection{Multi-modal Retrieval}
\label{sec:multimodal_results}
A key advantage of \textsc{ReFIT} is that it can operate independently of the choice of architecture for the bi-encoder and the cross-encoder, and is therefore not limited to working on only text input.
To demonstrate this, we apply our method to retrieval in a multi-modal setting. Specifically, we consider text-to-video retrieval, which involves retrieving videos that are relevant to a given textual query. 

The retriever and reranker for this experiment are based on BLIP \cite{li2022blip}, a state-of-the-art vision-language model that comprises two unimodal encoders and an image-grounded text encoder. 
The unimodal encoders encode image and text separately, akin to dual-encoders in text-to-text retrieval, and are trained with an Image-Text Contrastive (ITC) loss. The image-grounded text encoder injects visual information into the text encoder by incorporating a cross-attention layer, similar to a text-to-text cross-encoder, and is trained with an Image-Text Matching (ITM) loss. We refer the reader to \cite{li2022blip} for a more detailed description of BLIP's architecture and the pre-training objectives. BLIP can thus be used for retrieval with the unimodal encoders (which we refer to as BLIP$_{ITC}$), and for reranking with the image-grounded text encoder (which we refer to as BLIP$_{ITM}$). We use the output from BLIP$_{ITM}$ as the reranker distribution, which is then used to compute the distillation loss for updating the query representation that is output by BLIP$_{ITC}$. 

\begin{table}
    \centering
    \setlength{\tabcolsep}{1.0em}
    \def\arraystretch{1.2}
    \begin{tabular}{l|c|c}
    \hline
        \textbf{Method} & \textbf{R@10} & \textbf{R@100} \\
        \hline
        MIL-NCE & 32.4 & - \\
        VideoCLIP  & 30.0 & - \\
        FiT  & 51.6 & - \\
        \hdashline
        BLIP$_{ITC}$ & 69.0 & 92.1 \\
        Rerank (BLIP$_{ITM}$) & 74.6 & 92.3\\
        \textsc{ReFIT} & \textbf{74.7} & \textbf{92.9}* \\
        \hline
    \end{tabular}
    \caption{Recall of text-to-video retrieval methods on the MSRVTT benchmark. Rerank and \textsc{ReFIT} use BLIP$_{ITC}$ as the underlying retriever. Improvements with * are significant at $p<0.1$ as per the paired t-test.}
    \label{tab:multimodal}
    \vspace{-1.0em}
\end{table}
We evaluate using Recall@100 on the MSRVTT \cite{xu2016msr} text-to-video retrieval dataset, with BLIP$_{ITC}$ \cite{li2022blip} being our primary retrieval-only baseline along with other baselines taken from \cite{li2022blip}.
The \textit{Rerank} baseline uses BLIP$_{ITM}$ to rerank $K$=125 videos retrieved by BLIP$_{ITC}$.
Table \ref{tab:multimodal} compares performance on the 1k test split of MSRVTT. We see that BLIP$_{ITM}$
yields better ranking (as evident from higher Recall@10) than BLIP$_{ITC}$ as expected, but shows only minor gains in Recall@100. Crucially, \textsc{ReFIT} improves Recall@100 over the already strong BLIP$_{ITC}$ retriever, without a noticeable drop in Recall@10 compared to the BLIP$_{ITM}$ reranker.
\begin{table}[t]
    \centering
    \setlength{\tabcolsep}{1.0em}
    \def\arraystretch{1.2}
    \begin{tabular}{c|c|c|c|c}
    & \textbf{NQ} & \textbf{EntityQ} & \textbf{BEIR}& \textbf{Mr.TyDi} \\
    \hline
    Retrieve & 86.1 & 70.1 & 66.8 & 87.0  \\
    Rerank & 86.8 & 71.2 & 67.6 & 88.1  \\
    \textsc{TouR}$_{\text{hard}}$ & 87.0$^+$ & 71.9$^+$  & 68.4 & 88.7 \\
    \textsc{TouR}$_{\text{soft}}$ & 87.2$^+$ & 72.5$^+$ & 68.1 & 88.1\\
    \hline
    \textsc{ReFIT} & \textbf{87.6} & \textbf{72.6} & \textbf{69.2}* &  \textbf{89.7}* \\
    \hline
    \end{tabular}
    \caption{Recall@100 numbers for comparison of \textsc{ReFIT} with both variants of \textsc{TouR}. $^+$ corresponds to numbers directly taken from~\citet{sung2023optimizing}. Improvements with * are significant at $p<0.05$ according to the paired t-test.}
    \label{tab:tour_comparison}
    \vspace{-1em}
\end{table}

\subsection{Comparison with \textsc{TouR}}
\label{sec:tour_comparison}

In this section, we compare the performance of \textsc{ReFIT} with \textsc{TouR} on the passage retrieval benchmarks used in \citet{sung2023optimizing}, NQ~\cite{kwiatkowski2019natural} and EntityQuestions~\cite{sciavolino-etal-2021-simple} as well as the multidomain BEIR and multilingual Mr.TyDi benchmarks. For NQ  and EntityQuestions, we use the same retriever~\cite{karpukhin2020dense} and reranker~\cite{fajcik-etal-2021-r2-d2} as in \citet{sung2023optimizing}. The retriever and the reranker for BEIR and Mr.TyDi are the same as described in \S{\ref{sec:retriever_and_reranker}}. 
In Table \ref{tab:tour_comparison}, we can see that \textsc{ReFIT} consistently outperforms both \textsc{TouR} variants across various datasets. 
We believe that the lower performance of \textsc{TouR} can be attributed to its early stopping criterion for distillation updates. Specifically, \textsc{TouR} performs relevance feedback for 3 iterations, wherein in each iteration, distillation into the query vector continues until the top-1 retrieval result has the highest reranker score (for \textsc{TouR}$_{\text{soft}}$) or is a pseudo-positive (for \textsc{TouR}$_{\text{hard}}$). 
In contrast, \textsc{ReFIT} makes more distillation updates but for only one iteration (we do $n=100$ updates, which has been tuned). 
This makes it also considerably faster than \textsc{TouR}, as each additional iteration of relevance feedback in \textsc{TouR} comes with a high computational overhead (\S{\ref{sec:related-work}}). 
We show in \S{\ref{sec:multi_iterations}} that \textsc{ReFIT} can further benefit from multiple rounds of relevance feedback with continuous improvements over the course of three iterations.

\section{Discussion and Analysis}

This section describes additional experiments, providing further insights into \textsc{ReFIT}.

\subsection{Query vectors: the original and the new}
\label{sec:query-vectors-orig-new}

\begin{figure}[t]
    \centering
    \begin{subfigure}[c]{0.49\linewidth}
        \centering
     \includegraphics[width=1\linewidth]{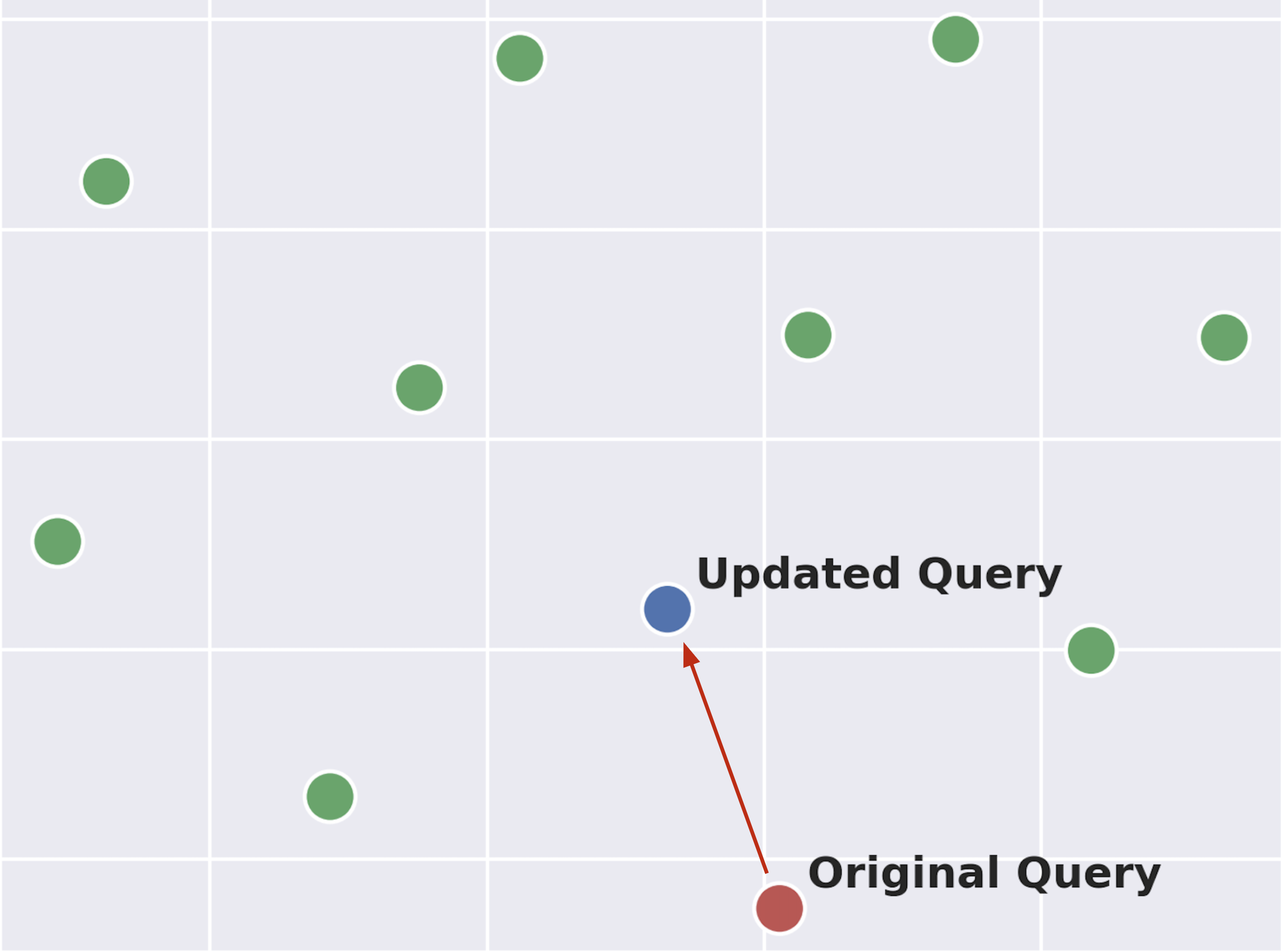}
     \label{fig:qvec_1}
     \vspace{-0.75em}
    \end{subfigure}
    \begin{subfigure}[c]{0.49\linewidth}
        \centering
     \includegraphics[width=1\linewidth]{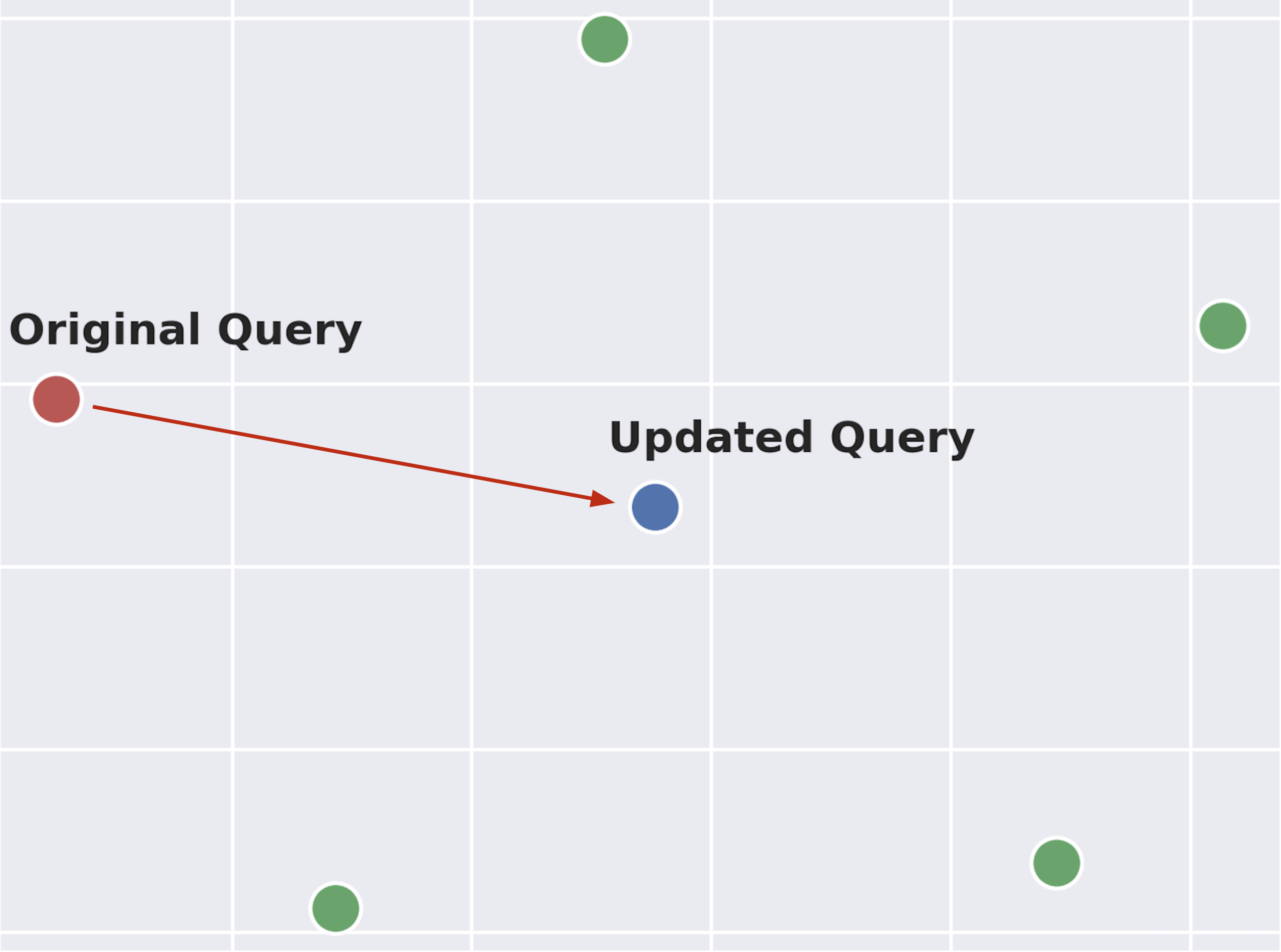}
     \label{fig:qvec_1}
    \vspace{-0.75em}
    \end{subfigure}
    \begin{subfigure}[c]{0.49\linewidth}
        \centering
     \includegraphics[width=1\linewidth]{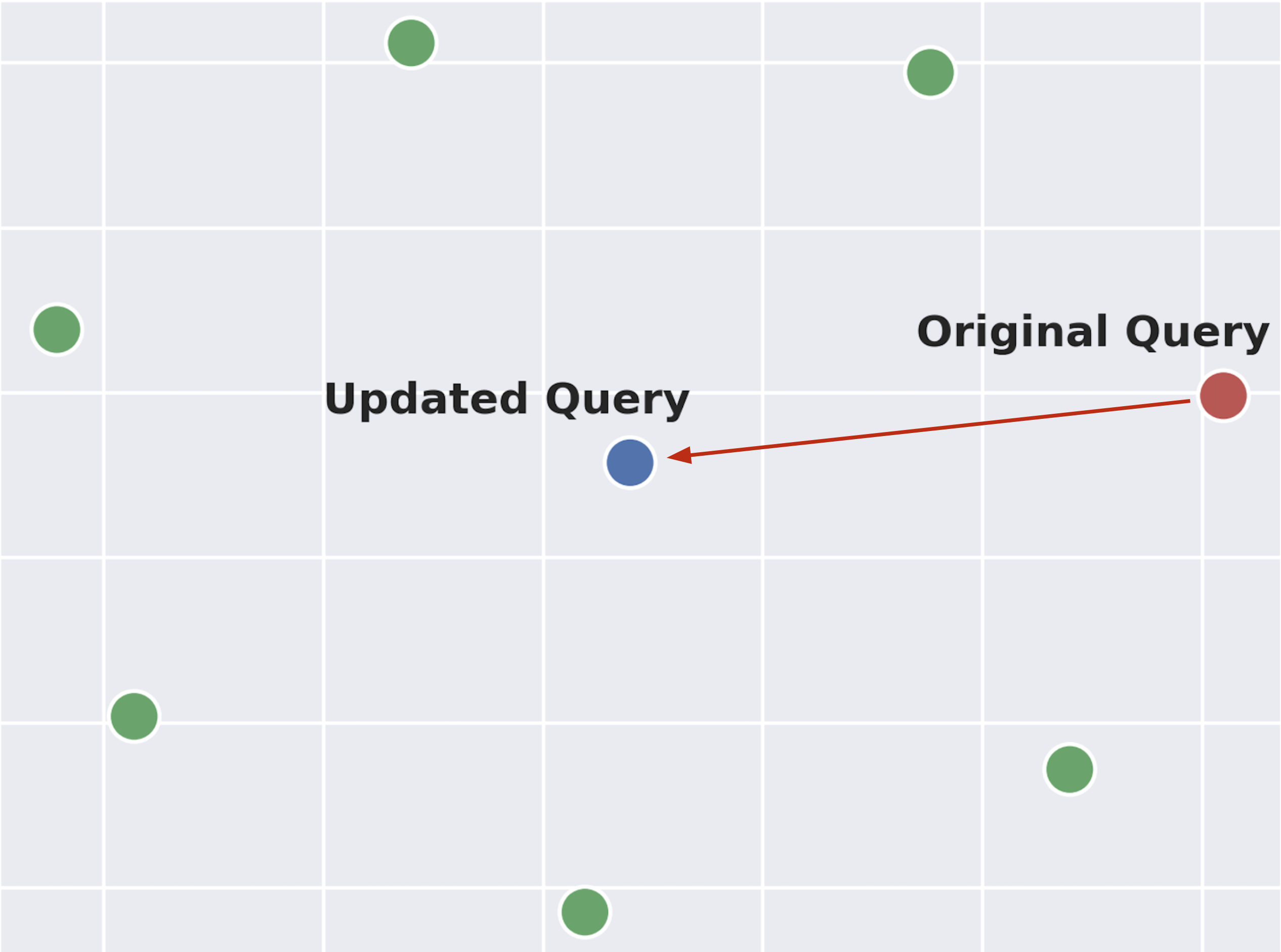}
     \label{fig:qvec_1}
     \vspace{-0.75em}
    \end{subfigure}
    \begin{subfigure}[c]{0.49\linewidth}
        \centering
     \includegraphics[width=1\linewidth]{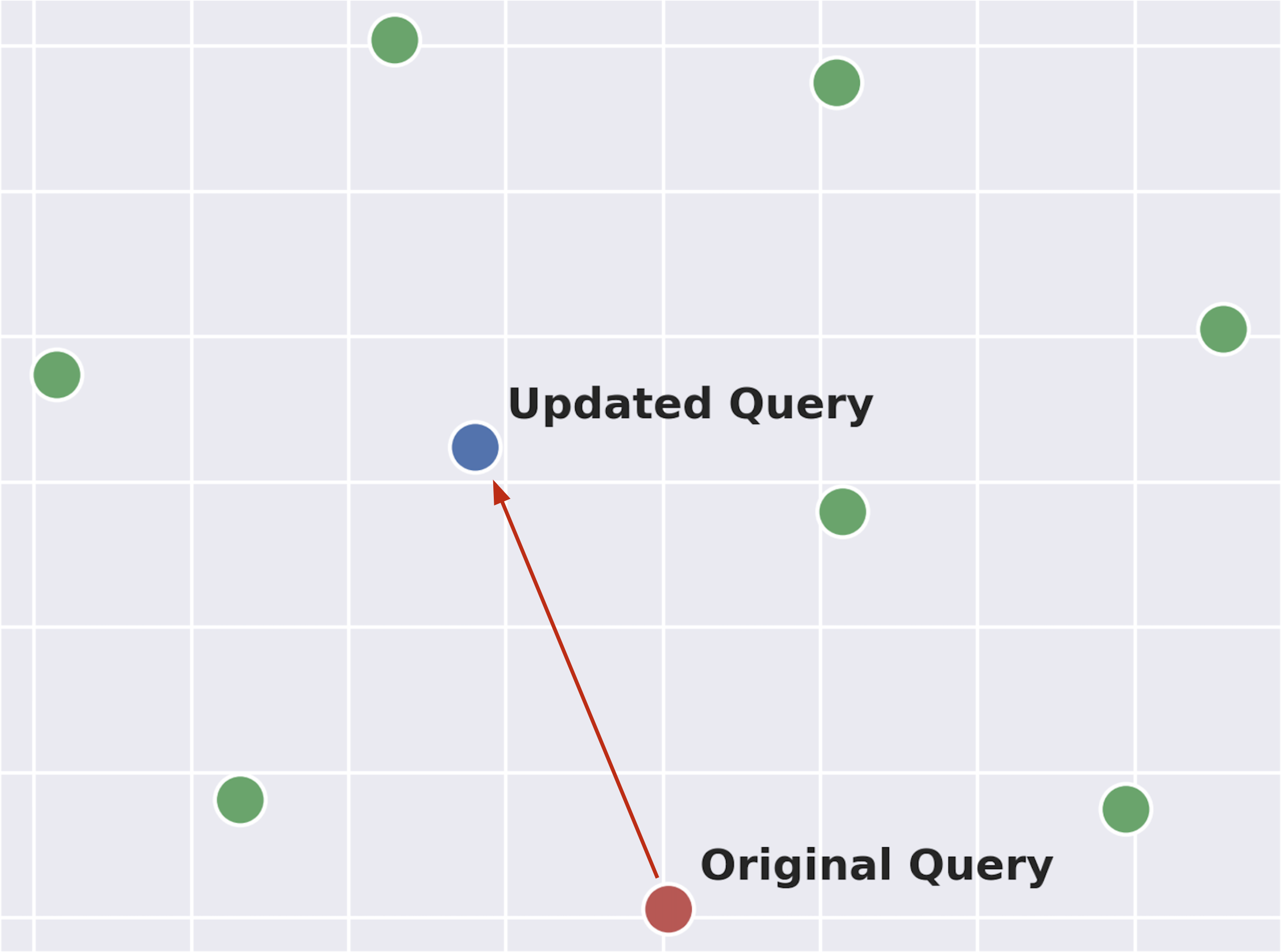}
     \label{fig:qvec_1}
    \vspace{-0.75em}
    \end{subfigure}
    \caption{t-SNE plots for some examples from BEIR, with the query vectors shown alongside the corresponding positive passages. The updated query vectors after \textsc{ReFIT} are now closer to the positive passages (in green).}
    \label{fig:TSNE}
    \vspace{-0.3em}
\end{figure}

To better understand how the updated query vector after reranker relevance feedback improves recall, we take a closer look at the query and passage vectors computed for a set of BEIR examples.
Figure \ref{fig:TSNE} shows t-SNE plots for four such examples, where each dot represents a vector, and the distance between any two points is their cosine distance.
As the figure clearly suggests, the reranker feedback brings the query vector in each case closer to the 
corresponding positive passage vectors, making the query align with an increased number of relevant passages and consequently improving recall.
Across different datasets in BEIR, we observed that the new query vector is also closer to the initially retrieved positives by 5-16\%.

We observe that the new positives discovered by the updated query vector are closest to a passage in the reranker's top 5 in 26\% of the cases (38\% for top 10; 55\% for top 20), confirming an effective transfer of the reranker's knowledge into the query vector.
Table \ref{tab:examples} provides some examples, showing how specific words and phrases 
in a passage within the reranker top-5 help retrieve additional candidates with lexical/semantic overlap (highlighted in green) via relevance feedback.
Interestingly, in the fourth example, an incorrect passage highly scored by the reranker leads to the subsequent retrieval of an actual positive candidate.

\begin{table*}[t]
    \small
    \centering
    \begin{tabular}{p{8em}|p{24em}|p{27em}}
    \multicolumn{1}{c|}{\textbf{Query}} & \multicolumn{1}{c|}{\textbf{Initial Retrieval (within Reranker Top-5)}}& \multicolumn{1}{c}{\textbf{Newly Retrieved Positive}} \\
    \hline
what mark of punctuation might indicate that more information is to come? & \textcolor{red}{A colon is the only punctuation mark that indicates that more information is to come}. \colorbox{lime}{An ellipsis}, which is used when you are quoting from another written source, \colorbox{lime}{indicates that something has been omitted}. & More answers. \textcolor{red}{A dash actually indicates more information to come.} \colorbox{lime}{An ellipsis is used usually to indicate that you're leaving out a} \colorbox{lime}{part} of a quote or other source. Parentheses or commas can also serve the same purpose, if used to insert an adjective clause.\\
    \hline
    treating tension headaches without medication & Most intermittent \colorbox{lime}{tension-type headaches} are easily treated with over-the-counter medications, including: 1 Aspirin. 2 Ibuprofen (Advil, Motrin IB, others) 3 Acetaminophen (Tylenol, others) & Instead of popping a pill when you get a headache, toss some almonds. For everyday \colorbox{lime}{tension-type headaches}, \textcolor{red}{almonds can be a natural remedy and a healthier alternative to other medicine}. \\
    \hline
    who drives the number 95 car in nascar & On October 2013, it was announced that McDowell would be moving to \colorbox{lime}{Leavine Family Racing's No. 95 Ford} for the 2014 \colorbox{lime}{NASCAR Sprint Cup Series} season. McDowell failed to qualify for the Daytona 500. & \textcolor{red}{Michael Christopher McDowell} is an American professional stock car racing driver. He currently competes full-time in the Monster Energy \colorbox{lime}{NASCAR Cup Series}, driving the \colorbox{lime}{No. 95 Chevrolet SS for Leavine Family Racing}. \\
    \hline
    who plays addison shepherd on grey's anatomy & In 2005, she was cast in her breakout role in the \colorbox{lime}{ABC series Grey's Anatomy, as Dr. Addison Montgomery}, the estranged spouse of Derek Shepherd. & \textcolor{red}{Kathleen Erin Walsh} is an American actress and businesswoman. Her roles include \colorbox{lime}{Dr. Addison Montgomery on the ABC} \colorbox{lime}{television dramas Grey's Anatomy} and Private Practice.\\
    \hline
    \end{tabular}
    \caption{Examples of how initial retrievals highly ranked (top-5) by the reranker (middle) helps retrieve new positives (right) via the updated query vector, due to important lexical and semantic overlap (highlighted in green). The text that contains the answer to the query is shown in red.} 
    \label{tab:examples}
\end{table*}

\begin{figure}[t]
    \centering
     \includegraphics[width=0.9\linewidth]{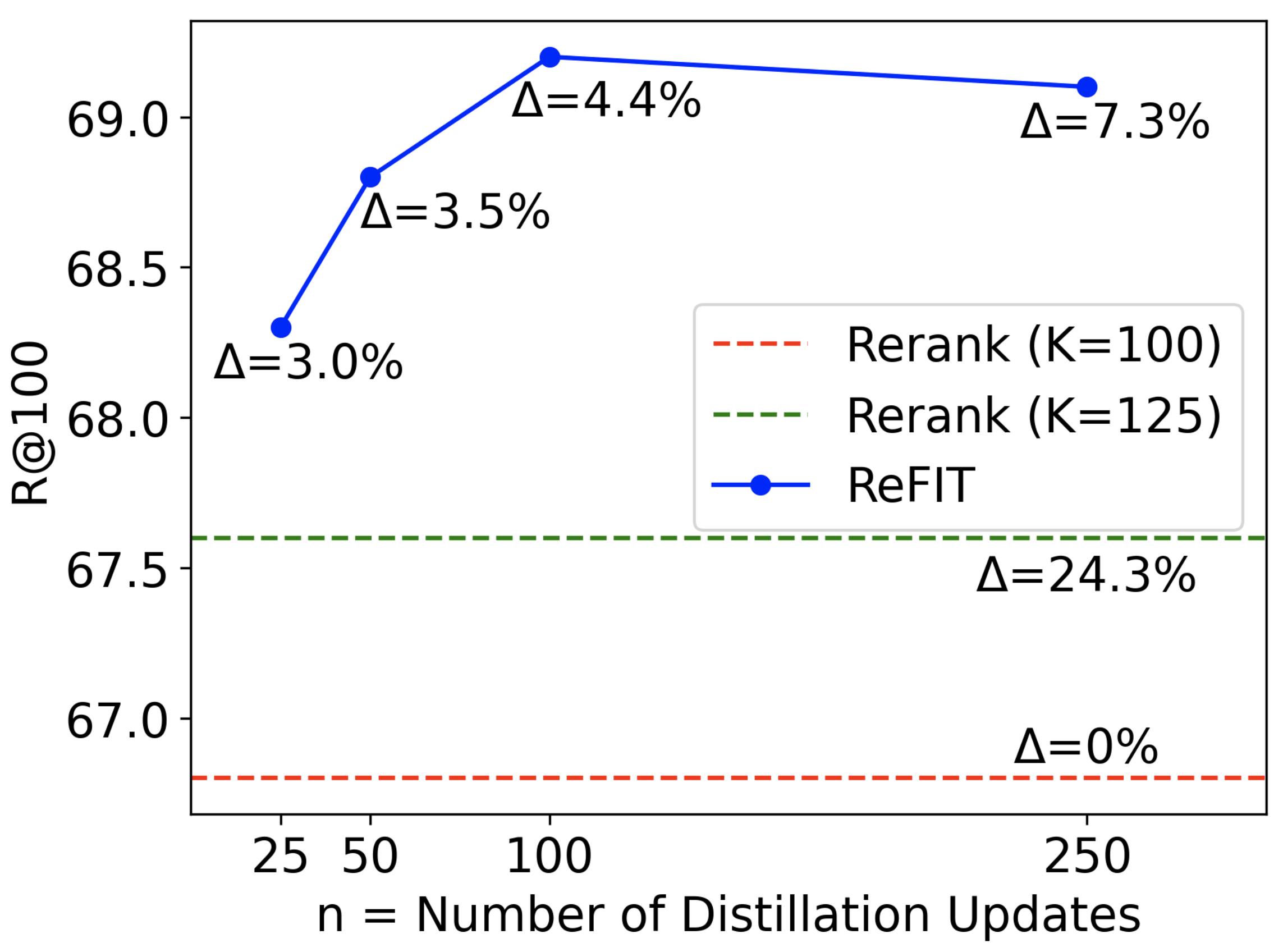}
    \caption{Plot showing the variation of \textsc{ReFIT} performance (R@100) with the number of distillation updates $n$ (where $\Delta$ is \% increase in latency on CPU).}
    \label{fig:n_variation}
    \vspace{-1em}
\end{figure}

\subsection{How much additional latency does our approach introduce?}
\label{sec:additional_latency}

Our proposed method introduces a distillation and an additional retrieval step into the standard R\&R framework. While retrieval takes constant time with respect to the number of updates $n$ in Algorithm \ref{alg4}, the latency of distillation is directly proportional to $n$. Figure \ref{fig:n_variation} demonstrates the effect of varying $n$ on both the latency and performance of our approach. 
The extra latency is computed with respect to a standard R\&R framework that runs with $K$=100.
With a mere 4.4\% increase in latency (for when $n$=100), our method produces a gain that is significantly larger than a more computationally expensive reranking of $K$=125 candidates which in turn corresponds to 24.3\% increase in latency on a CPU. Thereby, we demonstrate that, under latency constraints, our approach can be made faster by simply lowering the number of updates, while still surpassing the conventional strategy of reranking a larger pool of candidates for improving recall.

\subsection{How do smaller $K$ values affect results?}

\begin{figure}[t]
    \centering
     \includegraphics[width=0.9\linewidth]{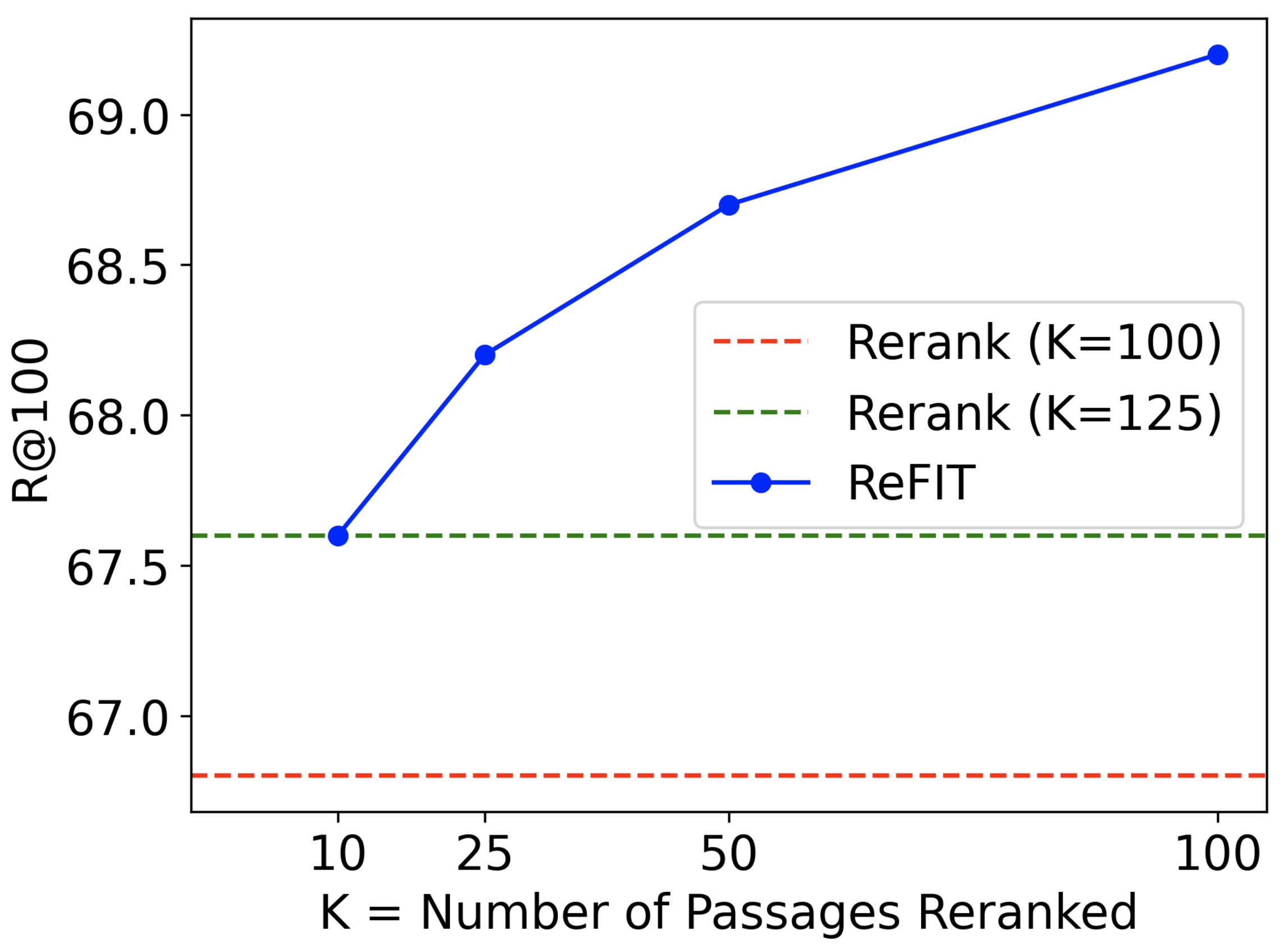}
    \caption{Plot showing the variation of \textsc{ReFIT} performance (R@100) with the number of reranked passages $K$ used for distillation supervision.}
    \label{fig:k_variation}
    \vspace{-1em}
\end{figure}

Our experiments described thus far are run in the standard setting of $K$=100: 100 passages are retrieved, reranked  and subsequently used to distill the reranker score distribution into the new query. Here we investigate how \textsc{ReFIT} performs as we vary $K$. Smaller values of $K$ correspond to a faster R\&R pipeline (as lower number of candidates are reranked), but it comes at the expense of the target teacher distribution now providing lesser supervision.
Figure \ref{fig:k_variation} shows Recall@100 of the post-relevance feedback retrieval step on BEIR for different values of $K$.
While a higher $K$ expectedly leads to a higher recall in general, we observe performance improvements over directly reranking 125 passages, even when considerably smaller number of passages are used for distillation.
Our approach can thus be easily tuned to achieve different accuracy-speed trade-offs depending on the requirements of the target application.


\subsection{\textsc{ReFIT} for multi-vector dense retrieval}
Our experiments thus far have been focused on single-vector dense retrieval, where queries and passages are encoded as individual vectors.
Multi-vector retrieval models like ColBERT~\cite{khattab2020colbert, santhanam-etal-2022-colbertv2}, on the other hand, compute token-level query and passage representations, subsequently employing a late-interaction mechanism for scoring. This section explores the application of \textsc{ReFIT} to multi-vector retrieval, specifically, ColBERTv2~\cite{santhanam-etal-2022-colbertv2}.
In this case, distillation (Step 3 in Figure \ref{fig:overall_framework}) updates embeddings of individual tokens in the query. We present results in Table \ref{tab:multi_vector} on a subset\footnote{Owing to the substantially larger index size inherent to multi-vector dense retrieval, we restrict this study to subsets of BEIR with $<100k$ passages in the retrieval corpus.} of the BEIR dataset, which clearly show that \textsc{ReFIT} can be effectively extended to multi-vector dense retrieval, as it consistently surpasses the performance of the ColBERTv2 retriever and outperforms the Rerank baseline (with $K=125$) in most cases. Notably, the training of ColBERTv2 by~\citet{santhanam-etal-2022-colbertv2} involved the use of a reranker's scores for supervision; our results in this section thus reinforce the finding of \S{\ref{sec:english_results}} that \textsc{ReFIT}'s inference-time distillation can be superior to ordinary knowledge distillation during training.


\begin{table}[t]
    \centering
    \setlength{\tabcolsep}{0.5em}
    \def\arraystretch{1.2}
    \begin{tabular}{c|ccc}
    \hline
     & \multicolumn{1}{c}{\textbf{ColBERTv2}}& \multicolumn{1}{c}{\textbf{Rerank}} & \multicolumn{1}{c}{\textbf{\textsc{ReFIT}}}   \\
    \hline
       NFCorpus   &  27.7 & 28.0 & \textbf{28.8}\\
       FiQA  & 62.8 &  63.7 &  \textbf{64.3}\\
       Scidocs & 35.8 & 36.6&  \textbf{38.5}\\
       Scifact   &  89.4 & \textbf{90.2} & 90.1\\
    \hline
    \end{tabular}
    \caption{Recall@100 (in \%) on a subset of the English BEIR benchmark, with Rerank and \textsc{ReFIT} using ColBERTv2 as the underlying retriever.}  
    \label{tab:multi_vector}
\end{table}


\subsection{Can multiple iterations of relevance feedback further improve results?}
\label{sec:multi_iterations}
Our relevance feedback approach improves recall when the updated query vector is used for a second retrieval step. 
Here we examine if further improvements are possible from more iterations of relevance feedback, i.e., running the following operations in a loop: (1) rerank the retrieval results from the previous iteration, (2) update the query vector via distillation from the reranker distribution, and (3) retrieve again. We note that this experiment operates under the assumption of a relaxed time budget, as the entire pipeline including the computationally expensive reranker must be executed $N$ times.
Table \ref{tab:multiple_rounds} shows performance on BEIR from $N$ iterations of relevance feedback, with $N=0$ corresponding to baseline retrieval (Contriever). We can see that recall improves with each additional round of relevance feedback; the biggest gain comes in the first round ($N=1$) and performance starts to saturate after $N=2$.

\begin{table}[!htb]
    \centering
    \setlength{\tabcolsep}{1.0em}
    \def\arraystretch{1.2}
    \begin{tabular}{c|c:ccc}
    \hline
     & \multicolumn{1}{c:}{\textbf{N=0}}& \multicolumn{1}{c}{\textbf{N=1}} & \multicolumn{1}{c}{\textbf{N=2}}  & \multicolumn{1}{c}{\textbf{N=3}}   \\
    \hline
     MS MARCO & 89.1 & 90.5 & 90.8 & \textbf{91.1}\\
       Trec-COVID & 40.7 & 51.5 & 54.5 & \textbf{55.8}\\
       NFCorpus   &  30.0 & 31.9& 31.7 & \textbf{32.0}\\
       NQ  &  92.5 &94.2 & 94.6 & \textbf{94.8}\\
       HotpotQA  & 77.7 & 80.4 & 81.4 & \textbf{81.9}\\
       FiQA  & 65.6 &  65.6& 66.3& \textbf{66.9}\\
       DBPedia  & 54.1 & 57.3& 58.7 & \textbf{58.8}\\
       Scidocs & 37.8 & 40.1& 40.0& \textbf{40.4}\\
       FEVER &  94.9 & 95.5 & 95.6& \textbf{95.7}\\
       Climate-FEVER  & 57.4 & \textbf{59.5} &59.3& 58.9\\
       Scifact   &  94.2 & \textbf{95.2} & 95.1& 93.7\\
       \hline
       Average &66.8 & 69.2 & 69.8& \textbf{70.0}\\
       \hline
    \end{tabular}
    \caption{Recall@100 (in \%) on the English BEIR benchmark after $N$ iterations of \textsc{ReFIT}. $N$=0 corresponds to the baseline Contriever model.}  
    \label{tab:multiple_rounds}
    \vspace{-1em}
\end{table}

\subsection{Further Discussion}
\subsubsection{The curious case of zero initial positives:} In \S{\ref{sec:query-vectors-orig-new}}, we presented an example where our method leverages a close negative among the initially retrieved candidates to later retrieve a positive passage. We find that in 24\% of the cases where the first-stage retriever retrieves no positive passages, our method can improve recall in a similar fashion. Among all cases where recall improves, however, 75\% have at least one positive in the top retrieved results. These results indicate that while the presence of positive candidates in the initial retrieval is useful, our relevance feedback approach can also generally leverage informative negatives to update the query vector in the right direction.

\subsubsection{Choice of Reranker:}
In the experiments comparing our approach to the R\&R framework, we used an efficient (yet high-performing) reranker both in the baseline model and as the teacher model for distillation. Would the results have been different if we used a more powerful (but computationally expensive) reranker instead?
To find an answer, it is essential to note that the final recall of an R\&R engine is inherently limited by the underlying retriever. For instance, the Recall@100 of an R\&R pipeline with $K$$=$$125$ cannot exceed the Recall@125 of the underlying retriever, irrespective of the quality of the reranker. The Recall@100 of \textsc{ReFIT} (BEIR: 69.2, Mr. TyDi: 89.7, MKQA: 68.2 and MSRVTT: 92.9) is consistently higher than the Recall@125 of the baseline retriever (BEIR: 68.9,  Mr. TyDi: 88.2, MKQA: 66.9 and MSRVTT: 92.8). These results clearly suggest that even the best reranker baseline would fail to attain the recall of our method. Further, we can safely expect a better reranker to improve the recall of \textsc{ReFIT} since leveraging a stronger teacher model (reranker) for distillation should lead to a better student (retriever query vector).

\section{Conclusion and Future Work}
We demonstrate that query representations can be improved using feedback from a cross-encoder reranker \textit{at inference time} for better performance of dual-encoder retrieval. This work proposes for distillation using relevance feedback from the reranker as a better and faster alternative to the traditional strategy of reranking a larger pool of candidates for improving recall.
\textsc{ReFIT} is lightweight and improves retrieval accuracy across different domains, languages and modalities over a state-of-the-art retrieve-and-rerank pipeline 
with comparable latency. Future work will focus on the potential integration of textual relevance feedback from large language models (LLMs). Additionally, a promising area of exploration lies in enhancing the interpretability by examining how relevance feedback influences the significance of individual query terms within the query representation.  

\section{Limitations}

\textsc{ReFIT} introduces an additional latency into a traditional retrieve-and-rerank framework.
The distillation time is only dependent on the number of updates, and is unaffected by the model architecture  and number of retrieved passages; the overall additional latency (30ms when $n$=100 for distillation plus 40ms for the second retrieval) amounts to an extra 17.5\% on GPU (or 4.4\% on CPU) when the number of retrieved passages $K$=100. However, it is noteworthy that \textsc{ReFIT} remains faster and exhibits superior performance compared to the standard approach of reranking a larger pool of candidates for improving recall.
Moreover, the efficacy of our approach is contingent upon the reranker providing a better ranking than the retriever. We anticipate that our method might provide minimal gains in situations where the performance of the retriever is close that of the reranker.

\clearpage

\appendix

\appendix

\section*{Appendix}

\section{Recall improves, but how good is the ranking?}
\label{sec:append_ranking_perf}
In \S{\ref{sec:results}}, we demonstrated that \textsc{ReFIT} improves Recall@100 over existing retrieval-only and reranking baselines. In many applications including web search, the ranks of the retrieved results matter too.
One important question for our method, therefore, is whether its output scores translate to an accurate ranking of the retrieved results; a failure to do so may necessitate a second expensive call to the reranker. We evaluate \textsc{ReFIT} for ranking on BEIR, Mr. TyDi and MKQA using the respective official ranking metrics. We note that MKQA contains judgments in the form of answer spans, and prior work only reports recall-based metrics. Hence, we use Recall@20 as a pseudo-measure of ranking quality in case of MKQA. Table \ref{tab:final_ranking_nums} shows the results: while Rerank expectedly yields big gains over Contriever, our approach is able to match Rerank on ranking performance. Combined with its improved recall of retrieval as reported in \S{\ref{sec:results}, this result indicates that our proposed method is suitable for both retrieval and ranking-oriented applications.




\begin{table*}[b]    
        \begin{subtable}[h]{0.49\linewidth}
        \small
            \centering
            \begin{tabular}{c|cc:ccc}
            \hline
            & \multicolumn{1}{c}{\textbf{BM25}}  & \multicolumn{1}{c:}{\textbf{ANCE}} & \multicolumn{1}{c}{\textbf{Contriever}} & \multicolumn{1}{c}{\textbf{Rerank}} & \multicolumn{1}{c}{\textbf{\textsc{ReFIT}}} \\
            \hline
             MS MARCO & 22.8 & 38.8  & 40.7  & 47.0  & 47.0 \\
               Trec-COVID  & 65.6  & 65.4     & 59.6  & 71.0  & 72.3 \\
               NFCorpus  & 32.5  & 23.7   & 32.8  & 34.5  & 35.1 \\
               NQ & 32.9 & 44.6   & 49.8 & 57.6  & 57.7 \\
               HotpotQA & 60.3   & 45.6  & 63.8   & 71.8 & 71.6 \\
               FiQA & 23.6  & 29.5   & 32.9  & 35.6  & 36.7 \\
               DBPedia  & 31.3 & 28.1   & 41.3  & 47.0  & 47.1 \\
               Scidocs & 15.8   & 12.2 & 16.5 & 17.0 & 17.1 \\
               FEVER & 75.3  & 66.9    & 75.8  & 81.9  & 81.9 \\
               Cli.-FEVER  & 21.3  & 19.8  & 23.7 & 25.5  & 25.7  \\
               Scifact  & 66.5  & 50.7  & 67.7  & 69.1 & 69.3 \\
               \hline
               Average & 40.7 & 38.7 & 45.9  & 50.7    & 51.0 \\
               \hline
            \end{tabular}
            \caption{nDCG@10 on the English BEIR benchmark.}
        \end{subtable}%
        \hfill
        \begin{subtable}[h]{0.49\linewidth}
        \small
            \centering
            \begin{tabular}{c|cc:ccc}
            \hline
              & \multicolumn{1}{c}{\textbf{mBERT}} & \multicolumn{1}{c:}{\textbf{XLM-R}} & \multicolumn{1}{c}{\textbf{Contriever}} & \multicolumn{1}{c}{\textbf{Rerank}} & \multicolumn{1}{c}{\textbf{\textsc{ReFIT}}} \\
            \hline
            Arabic   &  34.8  &  36.5   &  43.4  &  63.1   & 62.9  \\
            Bengali   &  35.1 & 41.7  &  42.3  &  62.4  & 62.9  \\
            English   & 25.7  & 23.0   & 27.1  & 37.7   & 37.9 \\
            Finnish   & 29.6 & 32.7  & 35.1   & 50.0  & 50.2 \\
            Indonesian  & 36.3  & 39.2  & 42.6  & 60.8    & 61.0\\
            Japanese  & 27.1  & 24.8  & 32.4  & 51.0   & 50.9 \\
            Korean   & 28.1   & 32.2   & 34.2 & 49.2  & 48.8 \\
            Russian & 30.0  &29.3  & 36.1 & 55.1   & 55.0 \\
            Swahili  & 37.4   &35.1  & 51.2  & 64.0  & 64.1 \\
            Telugu&  39.6  &54.7& 37.4 & 76.1 &76.1\\
            Thai& 20.3  &38.5 & 40.2  &  67.1 & 67.0 \\
            \hline
            Average & 31.3& 35.2   & 38.4   & 57.9 &  57.9  \\
            \hline
            \end{tabular}
            \caption{MRR@100 on the multilingual Mr.TyDi benchmark.}
        \end{subtable}%
        \vspace{0.5em}
        \begin{subtable}[h]{1.0\linewidth}
        \small
        \setlength{\tabcolsep}{1.0em}
            \centering
            \begin{tabular}{lcccccccccccccc}
            \hline
            & \textbf{avg} & \textbf{en} & \textbf{ar} & \textbf{fi} & \textbf{ja} & \textbf{ko} & \textbf{ru} & \textbf{es} & \textbf{sv} & \textbf{he} & \textbf{th} & \textbf{da} & \textbf{de} & \textbf{fr} \\ 
            \hline
            mBERT &45.3 & 65.5 &30.2 &38.9 &41.7 &34.5 &44.3 &52.4 &50.5 &32.6 &38.5 &52.5& 46.6 &53.8\\
            XLM-R &46.7 & 64.5 & 29.0 & 45.1 & 39.7 &34.9 &45.9 & 51.4 &56.1 &32.5 &49.4 &55.8 &48.3 &50.5\\
            \hdashline
            Contriever & 53.9 & 67.2 & 40.1 &55.1 &46.2 &41.7 &52.3 &59.3 &60.0 &45.6 &52.0 &62.0 &54.8 &59.3 \\
            Rerank & 59.6 & 70.7 & 47.0 & 59.9 & 54.6 & 49.1 & 59.3 & 64.7 & 65.5 & 53.2 & 57.9 & 66.1 & 61.5 & 64.3 \\
            \textsc{ReFIT} & 59.6 & 70.6 & 46.9 & 59.8 & 54.7 & 49.2 & 59.3 & 64.9 & 65.4 & 53.3 & 57.9 & 65.8 & 61.7 & 64.5 \\
            \hline
            \hline
            &  & \textbf{it} & \textbf{nl} & \textbf{pl} & \textbf{pt} & \textbf{hu} & \textbf{vi} & \textbf{ms} & \textbf{km} & \textbf{no} & \textbf{tr} & \textbf{cn} & \textbf{hk} & \textbf{tw} \\ 
            \hline
            mBERT & &52.1 &55.3 & 45.6& 49.5 &44.6 &46.9 &49.9 &21.5 &51.3 &42.7 &44.6 &45.3 &45.5\\
            XLM-R & & 45.4 &54.5 &48.5 &49.6 &47.3 &49.7 &54.0 &33.4 &53.7 &48.7 &42.4& 42.4 &42.0\\
           \hdashline
            Contriever & & 59.4 &60.9 &58.1 &56.9 &55.2 &55.9 &60.9 &26.2 &61.0 &56.7 &50.9 &51.9 &51.2 \\
            Rerank & & 64.9 & 66.1 & 62.8 & 63.0 & 59.9 & 62.6 & 65.4 & 30.5 & 65.9 & 61.9 & 57.8 & 59.2 & 57.5 \\
            \textsc{ReFIT} & & 64.8 & 66.0 & 63.0 & 63.4 & 59.7 & 62.3 & 65.3 & 30.7 & 65.3 & 62.1 & 60.0 & 58.3 & 57.5\\
            \hline
            \end{tabular}
            \caption{Recall@20 as a pseudo-measure of ranking for different approaches on the cross-lingual MKQA benchmark.}
        \end{subtable}
        \caption{Comparison of the ranking performance on the BEIR, Mr. TyDi and MKQA benchmarks.}
    \label{tab:final_ranking_nums}
\end{table*}

\section{Experiments with PRF baselines}
\label{sec:other_prf}
We also evaluated ANCE-PRF~\cite{yu2021improving} as a Pseudo-Relevance Feedback (PRF) baseline. ANCE-PRF trains a new PRF query encoder to output the query embedding by taking the query and top K (=5) documents retrieved by ANCE (concatenated with separator tokens) as input. In our experiments, we trained a PRF query encoder that takes the top K documents from our baseline retriever model, Contriever, as input. However, we found that the query embedding produced by the new PRF encoder did not improve upon the baseline Contriever model. This finding aligns with a previous study~\cite{li2022improving} that investigated the applicability of the ANCE-PRF method using dense retrievers other than ANCE. According to \citet{li2022improving}, improvement from the PRF query encoder was minimal once dense retrievers stronger than ANCE were considered. Further, \citet{li2022improving} considered dense retrievers~\cite{hofstatter2021efficiently} which were even weaker compared to the more powerful and recent Contriever model that we use as our retriever.

\begin{table}[!htb]
    \centering
    \small
    \begin{tabular}{c|c|c}
    \textbf{Domain} & \textbf{Published} & \textbf{Our Reproduction}\\
    \hline
        Trec-Covid &	67.5&	68.6\\
NFCorpus&	29.3&	29.4\\
NQ	&50.5	&50.2\\
HotpotQA	&53.3	&53.6\\
FiQA&	30.2	&30.4\\
DBPedia&	35.6&	35.8\\
Scidocs&	13.1&	12.8\\
FEVER	&67.6	&69.0\\
Climate-FEVER&	18.0	&19.1\\
Scifact&	56.8&	55.1\\
\hline
Average&	42.2&	42.4\\
\hline
    \end{tabular}
    \caption{nDCG@10 scores from our reproduction of RocketQAv2 compared those in prior work~\cite{santhanam-etal-2022-colbertv2, lin2023train}.}
    \label{tab:rocketqa_sanity}
\end{table}

\bibliographystyle{ACM-Reference-Format}
\bibliography{sample-base}


\end{document}